\let\origtime\time
\documentclass{jaa}
\usepackage{natbib}
\usepackage[colorlinks=true,citecolor=blue,bookmarks=false]{hyperref}
\usepackage{amsmath}
\let\time\origtime
\usepackage{graphicx}
\usepackage[skip=2pt,font=scriptsize]{caption}
\usepackage{caption}
\usepackage{subcaption}
\usepackage{graphicx}

\begin{document}\sloppy

%%paper title
%%For line breaks \\ can be used within title
\title{A global look into the world of interacting supernovae}

%%author names are separated by comma (,)
%%use \and before the last author name
%%use a * along with the number separated by comma
%% for the  author for correspondence
%%\textsuperscript{number} is used for affiliation
%%\affilOne, \affilTwo etc., upto \affilTwentyfive is possible
%%Please note the first letter after \affil is capitalised in the command
%%

\author{Anjasha Gangopadhyay\textsuperscript{1,2}, Kuntal Misra\textsuperscript{2}, Koji Kawabata\textsuperscript{1}, Raya Dastidar\textsuperscript{3} and Mridweeka Singh\textsuperscript{4}}
\affilOne{\textsuperscript{1}Hiroshima Astrophysical Science Center, Hiroshima University, 1-3-1 Kagamiyama, Higashi-Hiroshima, Hiroshima 739-8526, Japan\\}
\affilTwo{\textsuperscript{2}Aryabhatta Research Institute of observational sciencES (ARIES), Nainital 263001, India\\}
\affilThree{\textsuperscript{3}Millennium Institute of Astrophysics, Nuncio Monsenor Sótero Sanz 100, Providencia, Santiago, 8320000, Chile\\}
\affilFour{\textsuperscript{4}Indian Institute of Astrophysics, Koramangala, Bengaluru 560034, India}

%%escape two column mode for title, affiliation and abstract
%%by giving \twocolumn command as shown

\twocolumn[{

\maketitle

%%include \corres to print the corresponding author Email id
\corres{anjashagangopadhyay@gmail.com}

%%include \msinfo for
%%manuscript information such as
%%received, revised and accepted dates
%%
\msinfo{x xxxx xxxx}{x xxxx xxxx}

%%abstract
\begin{abstract}
Interacting supernovae (SNe) IIn and Ibn show narrow emission lines and have always been a mysterious and unsolved genre in SNe physics. We present a comprehensive analysis of the temporal and spectroscopic behaviour of a group of interacting SNe~IIn and Ibn. We choose SNe~2012ab, 2020cui, 2020rc and 2019uo as representative members of these SN sub-types to probe the nature of explosion. Our study reveals that SNe~IIn are heterogeneous, bright depicting multi-staged temporal evolution while SNe~Ibn are moreover homogeneous, comparatively fainter than SNe~IIn and short lived, but limited in sample to firmly constrain the homogeneity. The spectroscopic features display a great diversity in H$\alpha$ and He profiles for both SNe~IIn and Ibn. The representative SN~Ibn also show flash ionisation signatures of CIII and NIII. Modelling of H$\alpha$ reveals that SNe~IIn have in  general an asymmetric CSM which interacts with SN ejecta resulting in diversity in H$\alpha$ profiles. 
\end{abstract}

%%insert keywords separated by 3 hyphens using \keywords{words}

}]
%%close the twocolumn escape here

%%include \doinum{number}for the DOI number in the header
%%include \volnum{number} for the volume number in the header
%%include \year{yyyy} for  year of publication in the header
%%include \pgrange{num--num} page range of article in the header
%%include \artcitid{num} for the article citation id
%%include \lp to print last page of the article
%%include \setcounter{page}{pagenum} for the exact starting page of the article

\doinum{12.3456/s78910-011-012-3}
\artcitid{\#\#\#\#}
\volnum{000}
\year{0000}
\pgrange{1--}
\setcounter{page}{1}
\lp{1}

\section{Introduction}
Supernovae (SNe) are the giant stellar explosions that briefly outshine their host galaxy by radiating a huge amount of energy. SNe that have H in their outer spectra are classified as SNe~II whereas those that lack H features are classified as SNe~I. SNe that show evidence of strong shock interaction between their ejecta and pre-existing, slower circumstellar material (CSM) constitute an interesting, diverse, and yet poorly understood category of explosive transients commonly known as ``interacting SNe''. The progenitor star of these SNe become wildly unstable in time scale of years, decades, or centuries before explosion which makes it an interesting category. SNe IIn, with bright narrow Balmer lines of H in their spectra, were recognized as a distinct class of objects relatively late compared to other SN sub-types \citep{1990MNRAS.244..269S}. Related to this is a very similar sub-class of SNe known as SNe~Ibn, the identifying features of which are narrow He lines. The interpretations of ``interacting SNe" is very different from normal SNe as the ejecta is enshrouded by dense CSM and differs generally from that of "normal SNe" where emission comes from freely moving SN ejecta or radioactive decay at late times.

Interacting SNe result from CSM recombination following the ejecta-CSM interaction. At early times
these SNe may exhibit flash features \citep{2014AAS...22323502G}, i.e. narrow emission lines from highly ionised species (N V, He II and O V). In particular the spectra of SNe~Ibn typically show narrow emission lines of He having width between 1000-2000 km sec$^{-1}$; \citep{2007A&A...469L..31C} with a blue continuum. \cite{2017ApJ...836..158H} divided SNe Ibn typically into the P-cygni and emission subclass. The P-cygni subclass initially shows narrow P-cygni features of He which broadens over time while for the emission sub-class, the narrow He emission lines broaden with time during the evolution. SNe IIn spectra show  narrow-width (NW, $\sim$100 km sec$^{-1}$) components arising in the photoionised CSM \citep{1994ApJ...420..268C}, along with intermediate-width (IW, 1000-3000 km sec$^{-1}$) components due to either Thomson broadening of NW lines, or due to emission from gas shocked by the SN ejecta \citep{1994MNRAS.268..173C, 2009MNRAS.394...21D}. Some events also show very broad emission components arising from shocked ejecta (e.g. \citealt{1993MNRAS.262..128T}).  The light curves of SNe~IIn are diverse and heterogeneous \citep{2019ApJ...884...87T}, whereas SNe~Ibn light curves are homogeneous with decay rates consistent with  0.1 mag d$^{-1}$ (calcuated from peak upto 30 days post maximum light, \citealt{2017ApJ...836..158H}).

The intensity of the absorption and emission lines strongly depend on the density of the CSM and the velocity contrast between ejecta and CSM. If the density of the CSM is low, the emission from the CSM-ejecta interaction becomes visible only after the SN luminosity has faded, sometimes several years after the explosion. However, occasionally, the CSM near the SN is so dense that the CSM-ejecta interaction dominates the SN emission even at early phases. If the CSM is asymmetric, then the intensity of the emission and absorption strengths of the lines are visible depending not only on the CSM density but also on the viewing angle. The strength of the IW lines can be used to estimate the mass-loss rate of the progenitor during the pre-supernova evolution and is thus critical to estimate the ZAMS mass of the progenitor. While a Luminous Blue Variable progenitor (LBV) is predicted for some SNe IIn, some may originate from 8-10 M$_{\odot}$ stars undergoing core collapse as a result of electron capture after a brief phase of enhanced mass loss, or from more massive (25 M$_{\odot}$) progenitors, which experience substantial fallback of the metal-rich radioactive material \citep{2004MNRAS.352.1213C, 2012MNRAS.424..855K}. For SNe~Ibn, only two direct evidences (SNe~2006jc and 2019uo) have been reported so far suggesting Wolf-Rayet (WR) stars as their progenitors. However, late time UV/optical {\it HST} images have also proposed a binary progenitor scenario \citep{2007Natur.447..829P, 2021ApJ...907...99S}.

In this paper, we present the diversity in photometric and spectroscopic behaviour of a selected group of interacting SNe. We also study the asymmetries associated at the ejecta-CSM front of these SNe. The SNe are corrected for the Milky way and host galaxy extinction. The analysis is perford by adopting the luminosity distance for each SNe assuming $H_0 = 73$~km~s$^{-1}$~Mpc$^{-1}$.
The detailed temporal and spectral evolution of the interacting SNe is given in the following sections.

\section{Sample selection}
We select a heterogeneous sample of core collapse SNe to study the diversity in these events, particularly the interacting SNe. SNe~II comprises of ten objects:
SNe 1996al \citep{2016MNRAS.456.3296B}, 1998S \citep{2000MNRAS.318.1093F, 2001MNRAS.325..907F}, 1999em \citep{2003MNRAS.338..939E}, 2005ip \citep{2012ApJ...756..173S}, 2006gy \citep{2007ApJ...666.1116S}, 2007od \citep{2010ApJ...715..541A, 2011MNRAS.417..261I}, 2009kn \citep{2012MNRAS.424..855K},
2010jl \citep{2016MNRAS.456.2622J}, PS1-10adi \citep{2017NatAs...1..865K} and 2012ab \citep{2020MNRAS.499..129G}. 
%SNe 2006gy, PS1-10adi, and 2012ab are located close to the nuclei of host galaxies. 
We have in our sample highly interacting SNe~IIn 2006gy, 2010jl where prolonged interaction signatures are seen upto nearly a year post explosion along with shortly interacting SN 1998S, where the typical SN P-cygni profiles indicating ejecta signatures are visible a few months after the explosion. Transitional SNe from SNe IIL/IIP to IIn like 1996al, 2007od, 2009kn and SN IIP 1999em are also included for comparison.  Ten SNe comprise the sample of SNe~Ibn - SNe 2006jc \citep{2007Natur.447..829P}, OGLE12-006 \citep{2015MNRAS.449.1941P}, 2010al \citep{2015MNRAS.449.1921P}, 2011hw \citep{2015MNRAS.449.1921P}, 2014aki \citep{2017ApJ...836..158H}, 2014av \citep{2017ApJ...836..158H}, 2015G \citep{2017ApJ...836..158H}, 2015U \citep{2017ApJ...836..158H}, ASASSN-15ed \citep{2015MNRAS.453.3649P} and 2019uo \citep{2020ApJ...889..170G}. Some members are of P-cygni subclass and some are of emission subclass. Moreover, we have transitioning member (from SNe Ibn to Ib) like ASASSN-15ed in the sample. We choose SNe~2012ab and 2019uo as representative members of SNe IIn and Ibn respectively and show a detailed evolution highlighting the typical behaviour of these group. We also present new observations of SNe 2020cui and 2020rc from 3.6m Devasthal Optical Telescope (DOT) and examine the evolution of H$\alpha$ profile with respect to the sample.

\section{Temporal evolution}

\begin{figure*}[ht!]
	\begin{center}
		\includegraphics[scale=0.50]{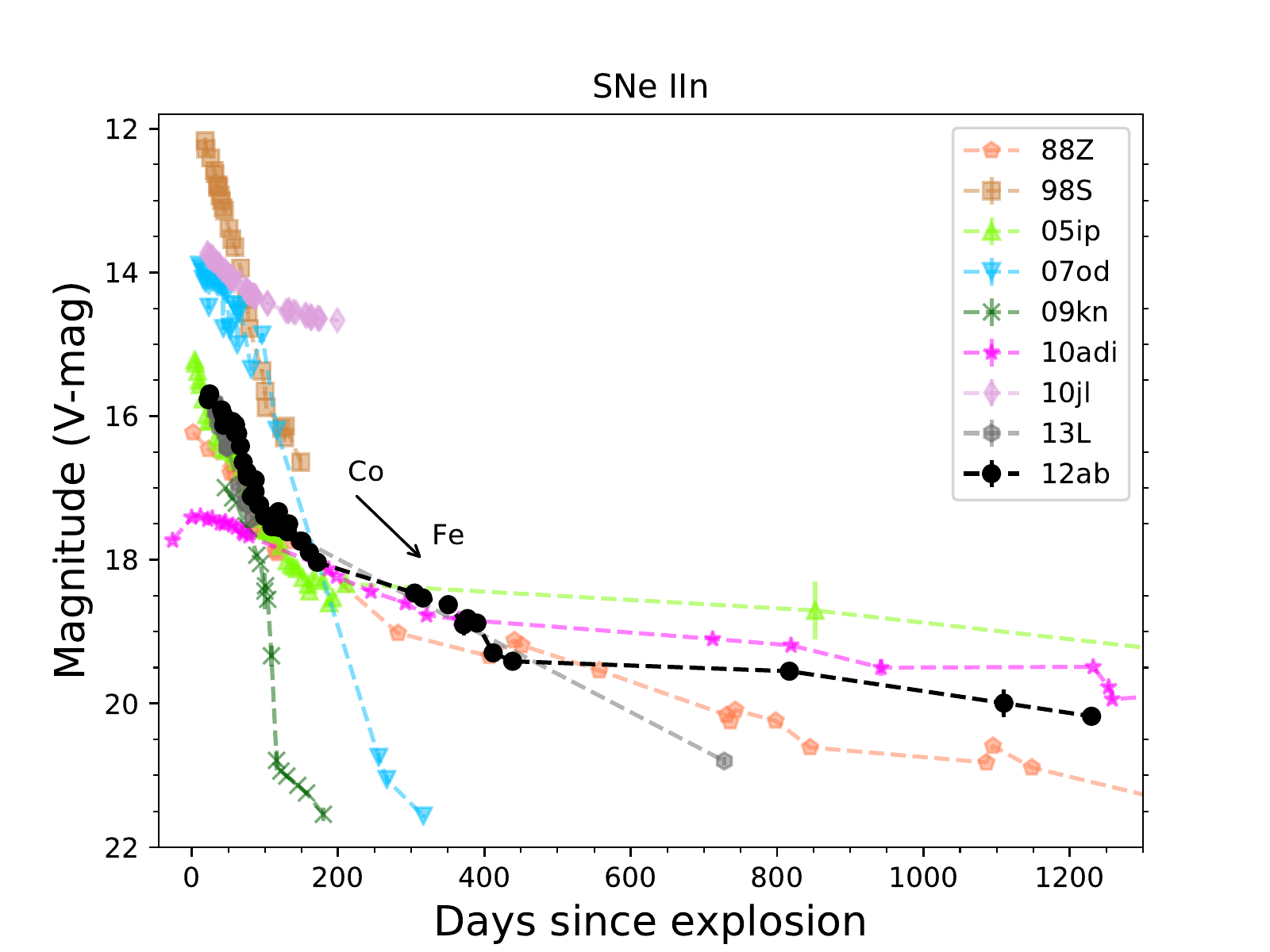}
		\includegraphics[scale=0.50]{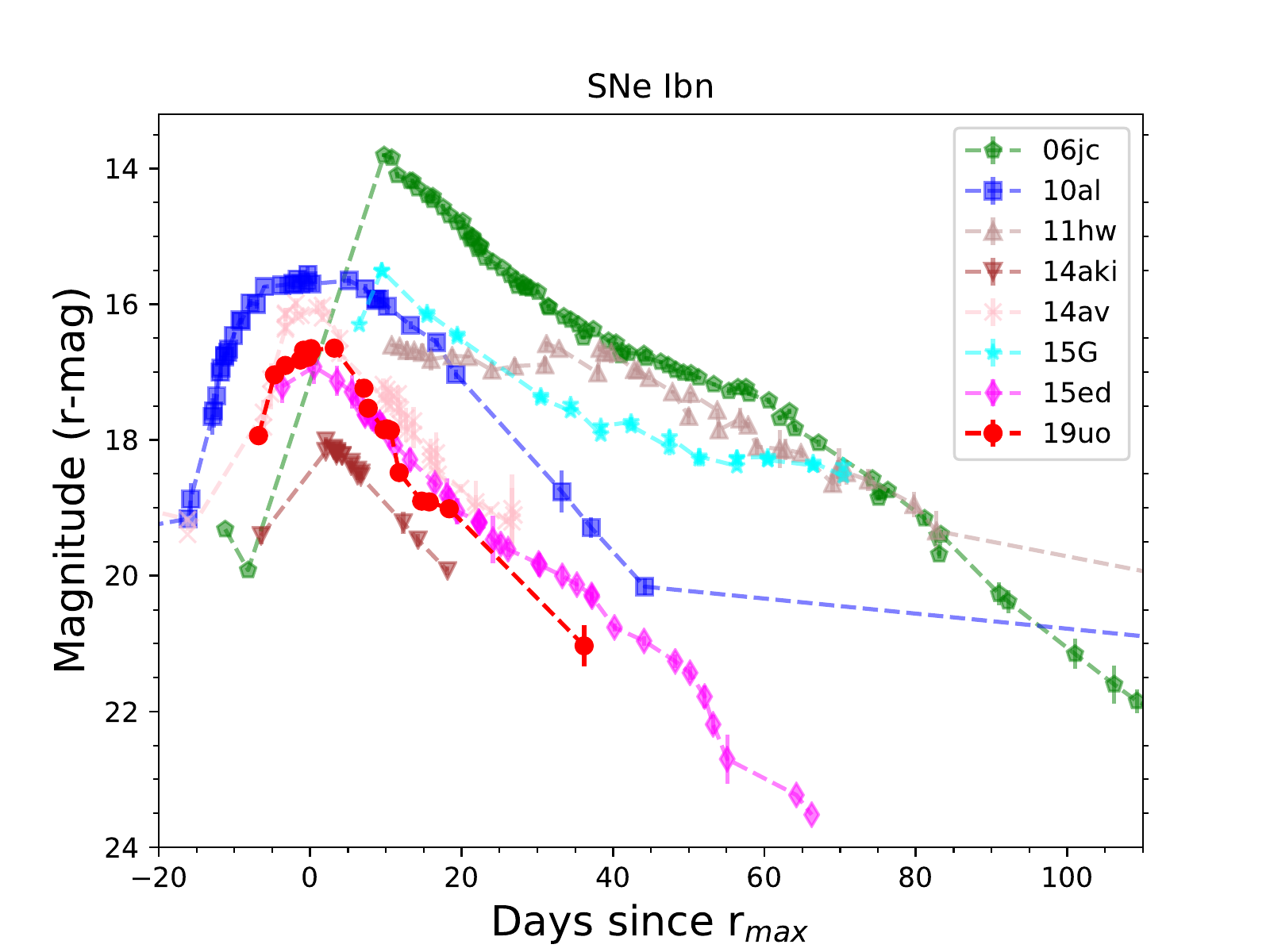}
	\end{center}
	\caption{{\bf Left:} The {\it V}-band light curve evolution of SN 2012ab along with other SNe~IIn. {\bf Right:} The {\it r}-band light curve evolution of SN 2019uo along with other SNe~ Ibn.}
	\label{fig:lc}
\end{figure*}

Figure~\ref{fig:lc} shows the light curve evolution of a group of SNe~IIn (V-band; left panel) and SNe~Ibn (r-band; right panel). The light curves of SNe~IIn are heterogeneous and long lived (beyond 200 days) where not only radioactivity plays a role, but also CSM interaction persists for a long time. On the contrary, the light curves of SNe~Ibn are homogeneous and short-lived typically upto 100-200 days post explosion. They decay with a typical rate of 0.1 mag days$^{-1}$ upto $\sim$ 30 days post maximum light \citep{2017ApJ...836..158H}. In this sample we highlight the high cadence light curves of SNe 2012ab (solid black circles; \citealt{2020MNRAS.499..129G}) and 2019uo (solid red circles; \citealt{2020ApJ...889..170G}). SN 2012ab was followed upto 1200 days after explosion displaying a rapid rise to maximum, a plateau lasting for about 2 months and a steep fall from the plateau. After 4 months the decline rate slows down below the $^{56}$Co decay rate upto 3.3 years post explosion indicating that not only radioactivity but CSM also plays an important role in the declining nature of SN~2012ab. The light curve evolution of SN 2019uo upto 50 days fades rapidly with a typical decay rate of 0.1 mag days$^{-1}$ indicating a short-lived CSM.

\begin{figure}[ht!]
	\begin{center}
		\includegraphics[width=\columnwidth]{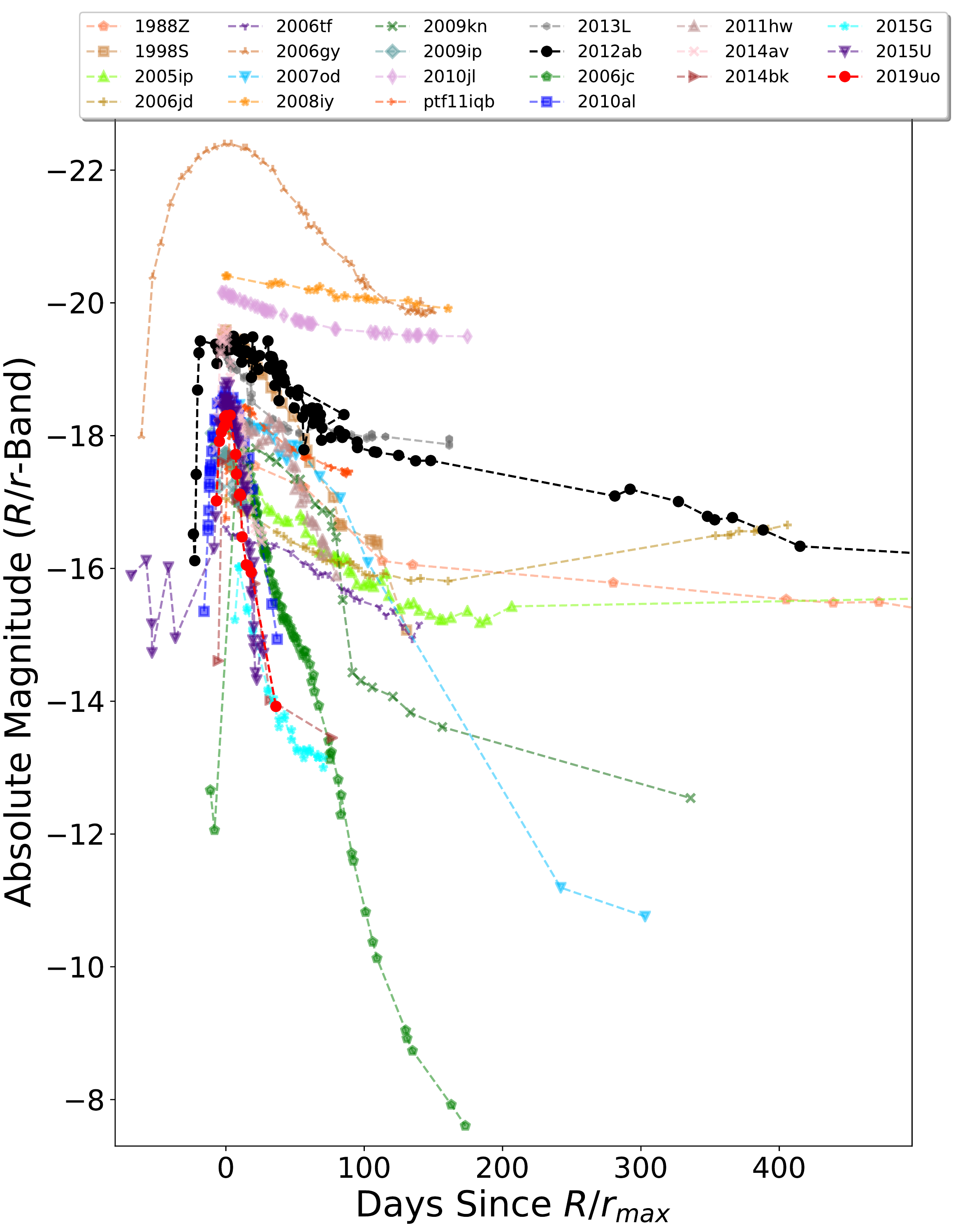}
	\end{center}
	\caption{Absolute magnitude light curves of a sample of SNe~IIn and SNe~Ibn. Solid black and red circles represent SNe 2012ab and 2019uo respectively.}
	\label{fig:abs}
	
\end{figure}

Figure~\ref{fig:abs} shows the absolute magnitude light curve evolution of a sample of SNe~IIn and SNe~Ibn. SNe~IIn shows a diversity in the brightness range varying between -22 to -18 mag while SNe~Ibn vary in brightness between -20 to -18 mag. 

\begin{figure*}[ht!]
	\begin{center}
		\includegraphics[scale=0.8]{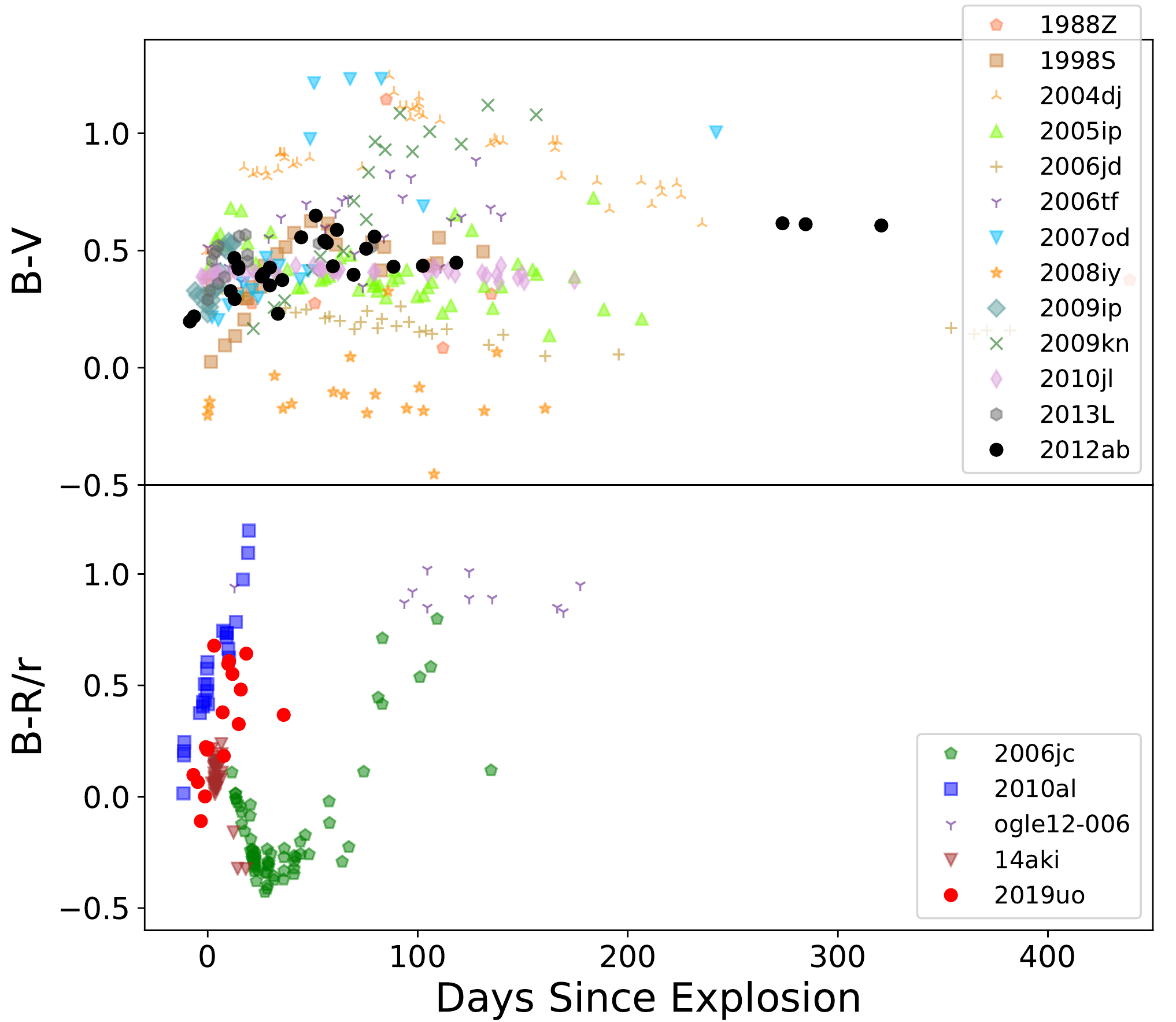}
	\end{center}
	\caption{{\bf Top:} (B-V) color evolution of SNe IIn. {\bf Bottom:} (B-R/r) color evolution of SNe Ibn. Solid black and red circles represent SNe~2012ab and 2019uo respectively.}
	\label{fig:color}
\end{figure*}

Figure~\ref{fig:color} shows the (B-V) and (B-R/r) color evolution of SNe IIn and SNe Ibn samples respectively. SN~2012ab (SN IIn) evolves redward very slowly from $\sim$ 0.1 mag to $\sim$ 0.7 mag between 22 days and 120 days.  This is expected because for interacting SNe, it is not only the SN ejecta which enhances the luminosity of the SN but CSM composition plays an important role in providing luminosity to the system and maintaining the blue colours. After 120 days and upto 440 days, the (B-V) color of SN 2012ab does not change significantly with a possible shallow decrease in (B-V) of $\sim$0.35 mag at intermediate epochs, in line with other interacting SNe, while normal SN II remain definitely redder \citep{2019A&A...631A...8H}. This is in agreement with most other strongly interacting SNe 1996al, 2005ip, 2006gy and 2010jl \citep{2016MNRAS.456.2622J} indicating that the color evolution is not only governed by the ejecta expansion as expected for normal SN 1999em and the weakly interacting SN 2007od that show a drift to the red, but, is rather dominated by CSM. SN 2009kn shows somewhat intermediate features between normal and interacting SNe. In SN 2019uo (SN Ibn) the (B-R/r) color increases up to 0.64 mag $\sim$20 days post explosion, subsequently becoming blue at $\sim$36 days. A similar trend is noticed in SN 2010al and iPTF14aki for which the (B-R/r) color increases up to $\sim$1 mag, $\sim$30 days post explosion. At similar epochs, the color evolution of SN 2006jc was extremely blue (-0.5 mag) and an overall flatter color evolution is noticed \citep{2007Natur.447..829P}. Thus, SNe~2019uo, 2010al and iPTF14aki shows a color evolution in which the early blue color are typical of SNe Ibn \citep{2016MNRAS.456..853P}. The transition to redder colors for SNe 2019uo and 2010al places their behavior between SNe Ib and most extreme SNe Ibn. 

\section{Spectral Evolution} 
\begin{figure*}[ht!]
	\begin{center}
		\includegraphics[scale=0.30]{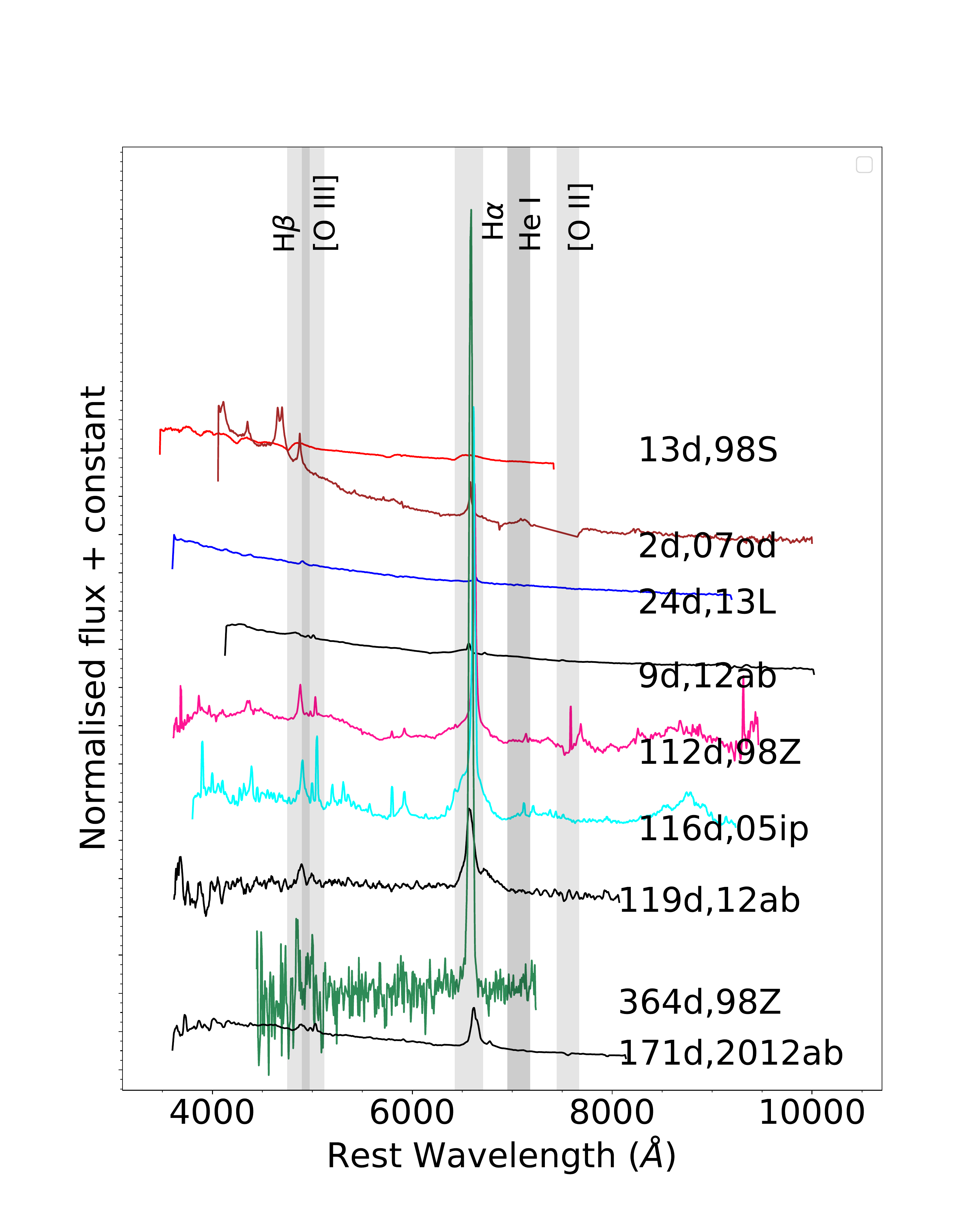}
		\includegraphics[scale=0.32]{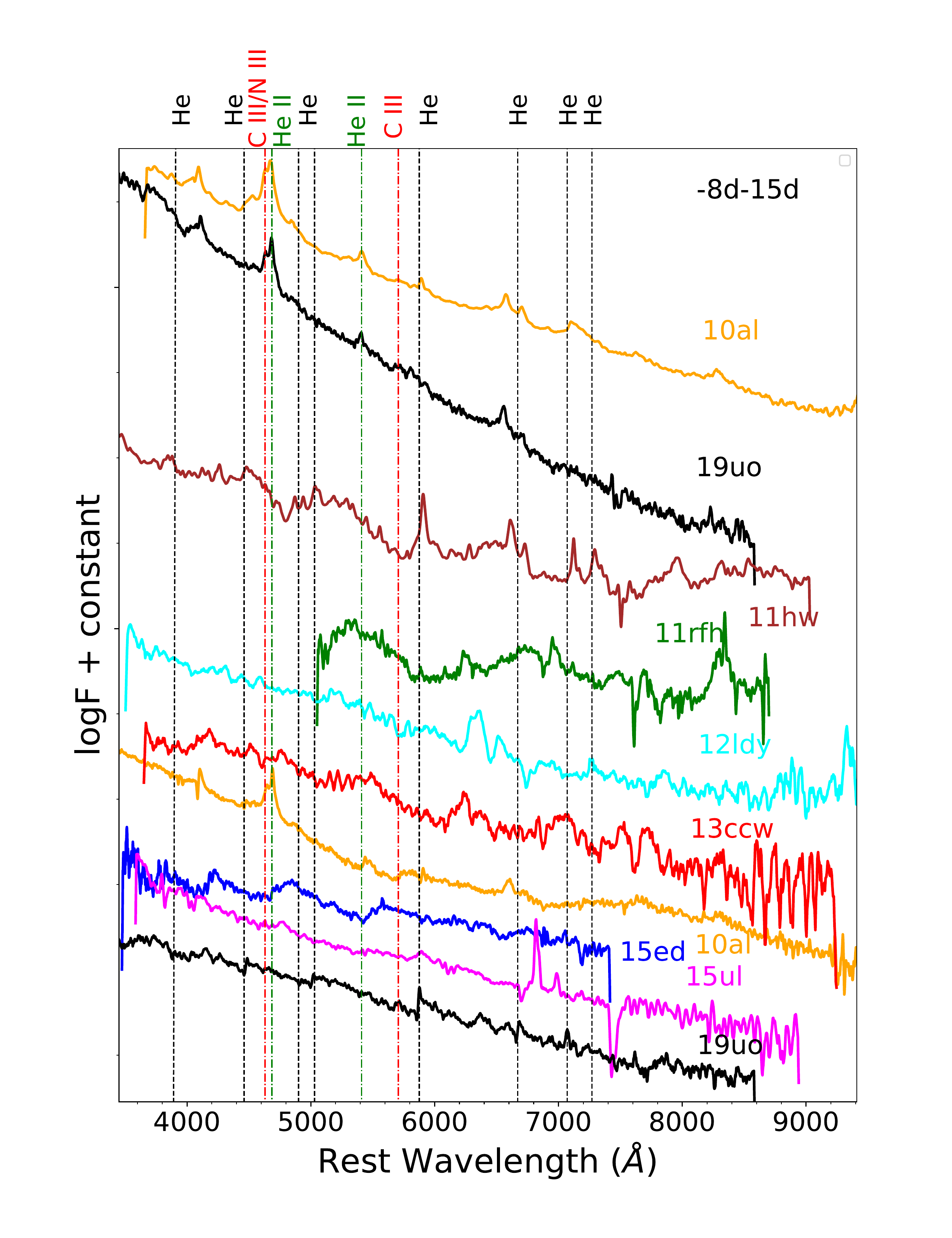}
	\end{center}
	\caption{Complete spectral evolution of a sample of SNe~IIn (left) and SNe~Ibn (right) along with the respresentative SNe 2012ab and 2019uo of each subclass.}
	\label{fig:spec}
\end{figure*}

\begin{figure}[t!]
	\begin{center}
		\includegraphics[scale=0.4]{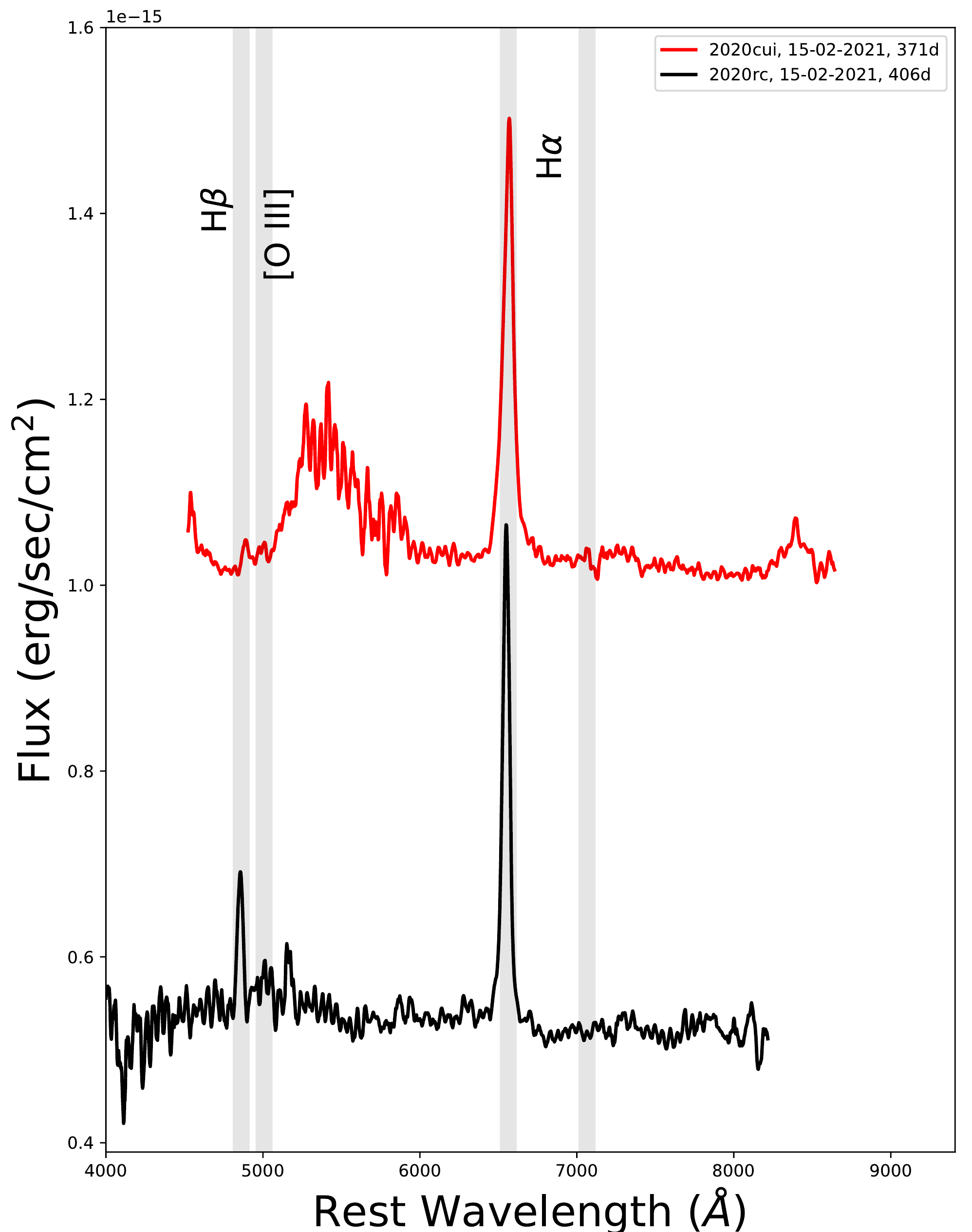}
	\end{center}
	\caption{The spectra of SNe~2020cui and 2020rc observed with the ADFOSC instrument mounted on the 3.6m DOT. The prominent lines are marked.}
	\label{fig:spec_dot}
\end{figure}

Figure~\ref{fig:spec} shows the early to late spectral evolution of SNe IIn (left) and SNe Ibn (right) samples highlighting SNe 2012ab and 2019uo. Prominent lines of Hydrogen i.e. H$\alpha$, H$\beta$ and H$\gamma$ are seen in SNe IIn. Even though the galaxy contamination lines are present, prominent lines of Mg I, Mg II and NaID start to develop with time. The Hydrogen features also develop and broaden over time. Prominent He I lines are seen in the spectral sequence of SNe Ibn. The early two spectra of SNe~2010al and 2019uo show a unique blue continuum and prominent emission features around $\sim$4660 \AA. We see a double peaked emission component, where the blue and the red components peak at 4643 \AA~ and 4682  \AA~ respectively. The blue component arises due to CIII and NIII while the red component arises due to He II. These signatures are interpreted as flash ionization signatures owing to the recombination of CSM. Flash ionization signatures of He II, C III and N III are also seen. These features are seen in very few SNe~Ibn and SNe~II. \cite{2010ATel.2491....1C} and \cite{2010CBET.2565....1S}  identified such lines to be originating from a Wolf-Rayet (WR) wind.

We observed two recent SNe~IIn 2020cui and 2020rc with the ARIES Devasthal Faint Object Spectrograph and Camera (ADFOSC)  \citep{2019arXiv190205857O} mounted on 3.6m Devasthal Optical Telescope (DOT), India with the aim to study the asymmetry at the ejecta-CSM front. The observations were performed at two epochs on 15-02-2021 and 15-03-2021. The data reduction was done using standard tasks in IRAF\footnote{Image Reduction and Analysis Facility} by extracting the one-dimensional spectrum and thereafter applying the wavelength and flux calibration. Figure~\ref{fig:spec_dot} shows the spectra of SNe 2020cui and 2020rc at a single epoch obtained on 15-02-2021. The spectra obtained on 15-03-2021 are noisy and contaminated by the host galaxy, hence we do not use them for further analysis.

\section{Asymmetries in H$\alpha$ profile}
\begin{figure*}[ht!]
	\begin{center}
		\includegraphics[scale=0.8]{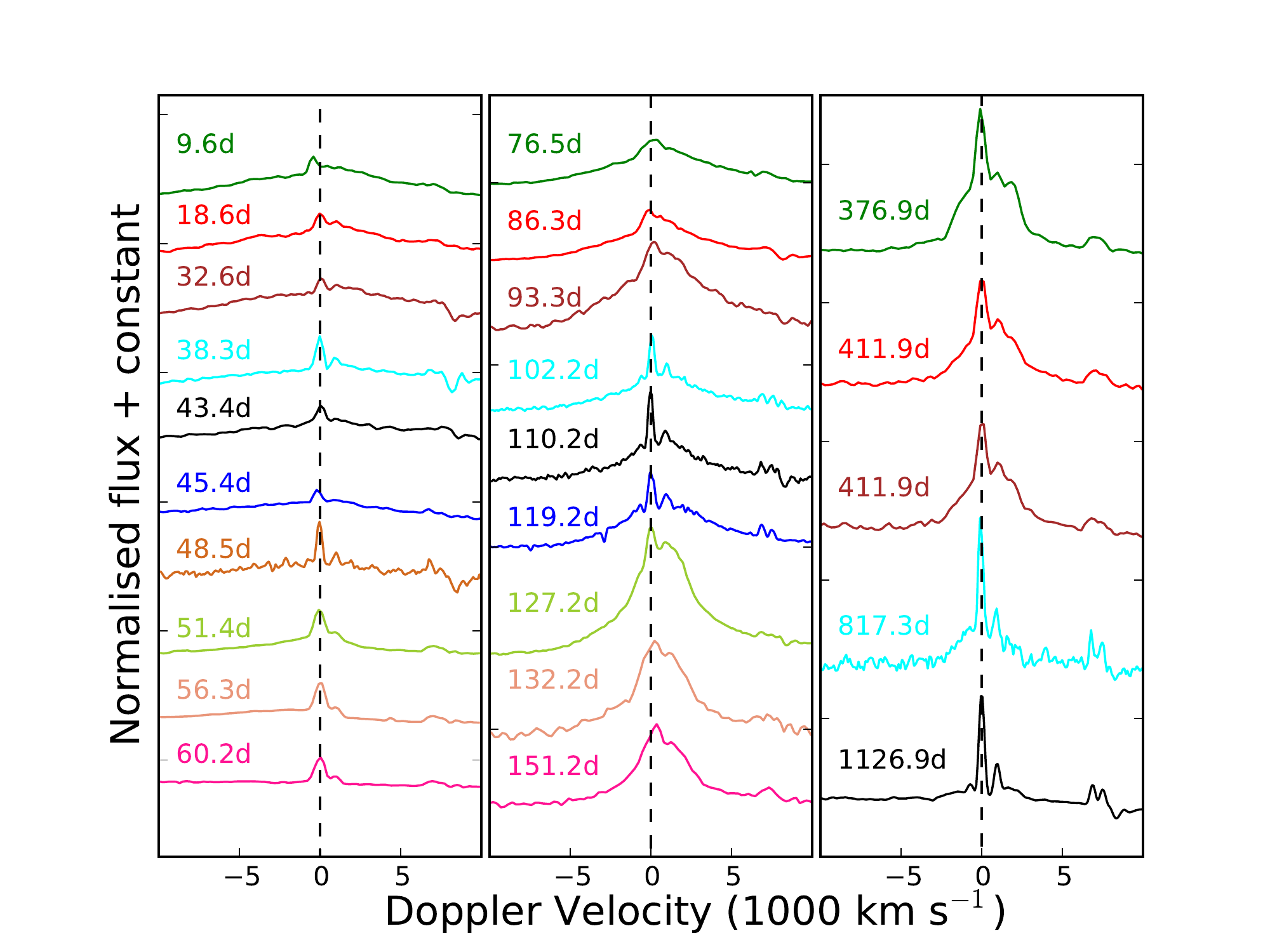}
	\end{center}
	\caption{H$\alpha$ evolution of a SN~IIn 2012ab that shows variation with time.}
	\label{fig:Halpha_evol_2012ab}
\end{figure*}

\begin{figure}[]
	\begin{center}
		\includegraphics[width=\columnwidth]{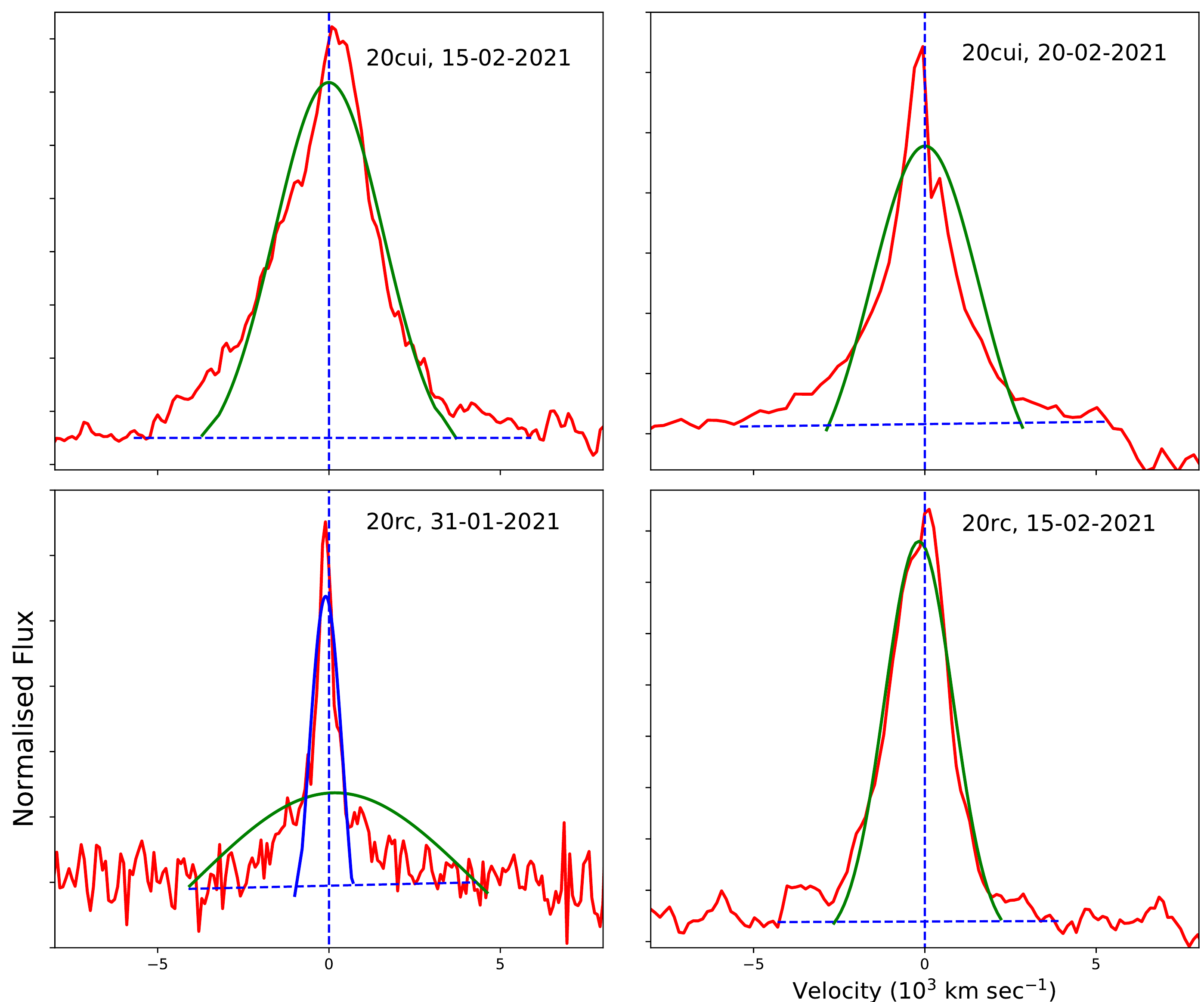}
	\end{center}
	\caption{Evolution of H$\alpha$ profile in the spectra of SNe~IIn 2020cui and 2020rc.}
	\label{fig:Halpha_evol}
\end{figure}

During late times (1-2 years post explosion), the fast SN ejecta may be heated predominantly by radiation from the CSM interaction shock that propagates backward into the outermost ejecta, creating a very different time dependent spectrum than is seen in a normal SN heated from the inside out by radioactive decay  \citep{2017hsn..book..403S}. The above scenario is true, provided we have a perfectly symmetric ejecta-CSM configuration. If the geometry is asymmetric, any of the zones in ejecta-CSM interaction can be seen simultaneously and at different characteristic velocities, potentially making the interpretation quite complicated \citep{2017MNRAS.466.3021S}. Absorption features may or may not be seen, depending on the viewing angle. Thus, for the time being, unraveling the asymmetries in the spectral evolution of interacting SNe is more like an art form. The probe is of course the H$\alpha$ profile which shows huge variations. Figure~\ref{fig:Halpha_evol_2012ab} shows the H$\alpha$ line evolution of SN~2012ab, as an example, where we see an evolution of H$\alpha$ from 9.6 days to 1126.9 days post explosion. From 9.6 days to 45.4 days, we see a narrow H$\alpha$ component with full width at half maximum (FWHM) velocities $\leq$ 1000 km sec$^{-1}$. In these spectra, a broad P-Cygni component is seen which is signature of the SN ejecta in addition to the narrow residual component indicative of pre-existing CSM. From 45 days to about 76 days, the H$\alpha$ component grows in strength with velocities ranging from 8,000 -- 18,000 km sec$^{-1}$. This shows an enhanced interaction of SN ejecta with CSM and gradual diminishing of SN features. Starting on 76.5 days, the intermediate width component also diminishes in FWHM velocities. There is, therefore a clear indication of interaction of very fast ejecta ($\sim$25,000 km sec$^{-1}$ red terminal velocity) with a receding blob of CSM. This continues upto 800 days where the redshifted part of the ejecta is completely engulfed by the CSM and we see a perfectly symmetric structure in the 1100 days.

\begin{figure}[]
	\begin{center}
		\includegraphics[width=\columnwidth]{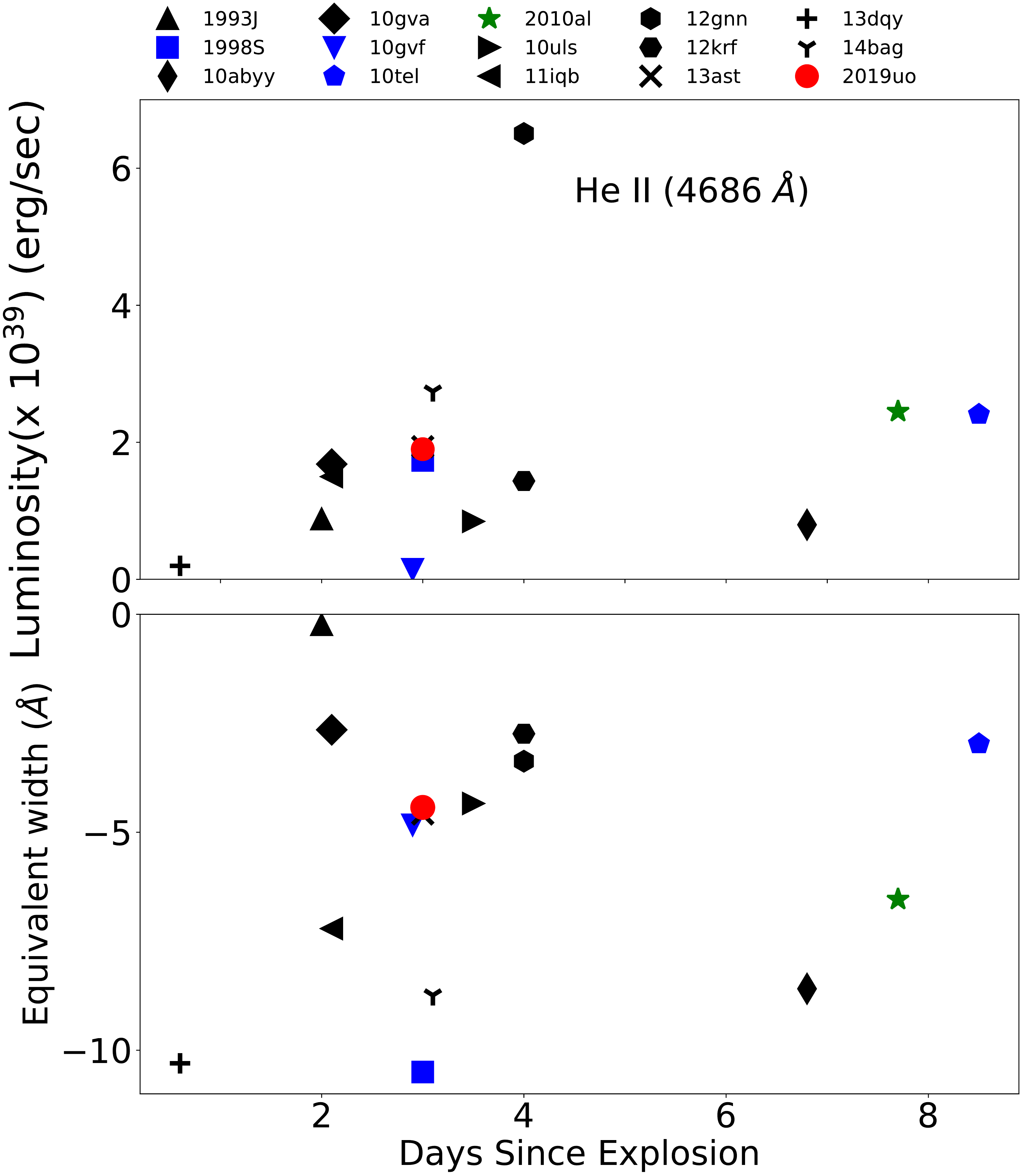}
	\end{center}
	\caption{He II luminosity and equivalent width estimation for a sample of SNe~Ibn. Blue symbols: SNe IIn, Black symbols: SNe~II (IIb, IIP and IIL), Red symbol: SN~2019uo, Green symbol: SN~2010al}
	\label{fig:lum_eqw}
\end{figure}

Figure~\ref{fig:Halpha_evol} shows the H$\alpha$ evolution of two SNe~IIn 2020cui and 2020rc. For a comparison we take spectra observed on 20-02-2021 (SN 2020cui) and 31-01-2021 (SN 2020rc) from WISERep (https://www.wiserep.org/). We modelled the H$\alpha$ profile of both SNe 2020cui and 2020rc using Gaussian/Lorentzian profiles. Defining and choosing a continuum is very critical for the required analysis. We have selected the continuum in a region external to the line region by at least 50 \AA. The selection of continuum and consistency of fits were repeated and checked several times. The first spectrum of SN~2020cui is fit by a Gaussian profile of intermediate width FWHM 3088 km sec$^{-1}$. In the second spectrum, we were able to fit a Gaussian with FWHM velocity of $\sim$3977 km sec$^{-1}$. The intermediate width lines of H$\alpha$ suggests ongoing interaction. We attempted to fit the 15-03-2021 spectrum of SN 2020cui after removing the host galaxy contribution and find that even though the H$\alpha$ is fit by a narrow component with velocity $\leq$300 km sec$^{-1}$, the spectrum is very noisy with a poor signal to noise ratio and the H$\alpha$ line cannot be resolved, given the limitations of the instrument. %The physical interpretation behind this scenario could possibly be that the SN initially shows strong interaction with a dense CSM in the month of February while in March the interaction has ceased and the narrow H$\alpha$ seen is seen only due to the residual host galaxy contamination.

In SN 2020rc, the 31-01-2021 profile is well fit with a narrow and a broad component of FWHMs of 980 km sec$^{-1}$ and 7800 km sec$^{-1}$ respectively. In the 15-02-2021 spectrum, the broad emission is gone and is completely overtaken by the intermediate width Gaussian profile of FWHM velocity $\sim$2000 km sec$^{-1}$. No emission from SN or host is seen in the 15-03-2021 spectrum likely due to the observational limitations of a 4m class telescope. In SN 2020rc an initial narrow component originating due to pre-shocked CSM and a broad component indicative of SN ejecta signatures is present. However, as time proceeds the ejecta is completely overtaken by the CSM and we see intermediate width features indicative of strong interactions. \cite{2017ApJ...836..158H} comment that viewing angle plays an important role in the determination of the line profiles of Hydrogen and Helium for SNe~Ibn. So, the narrow emission component is seen because the observer is located face on, while a P-cygni feature could have been traced if the observer is located edge on. 

To ascertain the nature and progenitor of SNe~Ibn, we estimated the luminosities and equivalent widths of a group of SNe~II (IIP, IIn, IIb) from a sample of \cite{2016ApJ...818....3K} and combined them with two SNe~Ibn with signatures of flash ionisation. Since H line are typically not present in SNe~Ibn and are contaminated by host galaxy lines, we choose relatively isolated line of  He II 4686 \AA. Since the He II lines are much narrower than lines from the SN ejecta, they can serve as a good tool for probing the flash-ionized CSM. When measuring the luminosities, we removed the continuum by fitting a linear function. Figure~\ref{fig:lum_eqw} shows the luminosity and equivalent width of SNe~II with the representative SN~2019uo. The luminosity is quite similar to SN~1998S. Since the flash ionised lines are tracers of WR signatures, higher luminosity may hint to more massive progenitor star. Similar conclusions can be inferred for the equivalent width estimations. However, given the circumstellar environment condition, these two are not that much evident to firmly comment on the nature of the progenitor.

%%Use table environment for a table in one column

%\begin{table}[htb]
%% use tabular font for a smaller size font
\tabularfont
%\caption{Table fitting in a single column.}\label{tableExample} %%10/12
%\begin{tabular}{lccccc}
%\topline
%one& two &three&four&five&six\\\midline
%1&2&3&4&5&6\\
%aaa&bbbb& ccccc&dddd&eeeee&ffffff\\
%\hline
%\end{tabular}
%%use \tablenotes{footnote} to get the table foot note
%\tablenotes{Sample table footnote}%%9/11
%\end{table}

%%Use table* environment to get the table spanning both the columns

%\begin{table*}[htb]
%\tabularfont
%\caption{Caption text here}\label{secondTable}
%\begin{tabular}{lccccccccccccr}
%\topline
%\textbf{head1}&\multicolumn{11}{c}{\textbf{head2}}&\textbf{head3}\\
%\midline
%one& two &three&four&five&six&seven&eight&nine&ten&eleven&twelve&thirteen\\
%1&2&3&4&5&6&7&8&9&10&11&12&13\\
%aaa&bbbb&cccc&ddddd&eee&ffff&ggggg&hhhhhhhh&iiii&kkkkkk&hhh&jjjjjj&lllll\\
%\hline
%\end{tabular}
%\tablenotes{Table footnote here. Table spanning both the columns.}
%\end{table*}

%%An example of a figure

%\begin{figure}[!t]
%\includegraphics[width=.8\columnwidth]{fig1.eps}
%\caption{caption goes here}\label{figOne}
%\end{figure}

%%An example of a double column figure
%%Use figure* environment

%\begin{figure*}
%\centering\includegraphics[height=.15\textheight]{fig1.eps}
%\caption{caption spanning two columns}
%\centering\includegraphics[height=.25\textheight]{fig1.eps}
%\caption{caption here}
%\end{figure*}

\vspace{-2em}
\section{Conclusion}
The paper summarises the temporal and spectroscopic evolution of a sample of SNe~IIn and Ibn which typically show narrow emission lines of H and He owing to the interaction of the SN ejecta with the dense CSM. The light curves of SNe~IIn shows a heterogeneous evolution with varied decay rates and late time flattening mostly due to CSM interaction. The light curves of SNe~Ibn, in contrary, are homogeneous with a typical decay rate of 0.1 mag/day and are short lived. On absolute brightness scale, the SNe~IIn are brighter reaching upto -22 mag while SNe~Ibn are relatively fainter reaching upto -20 mag. This may indicate a more dense CSM for SNe~IIn showing significant interaction. The color curves become red after a long time which suggests a diverse variation for SNe~IIn owing to long term CSM. We show the spectroscopic evolution of SNe~IIn and Ibn highlighting SNe 2012ab and 2019uo. Prominent H$\alpha$ and He lines are traced for both the events. SN~2019uo also shows signature of flash ionisation owing to the recombination of CSM front. SN~2012ab shows a great variation in H$\alpha$ profile showing narrow feature from pre-shocked CSM, intermediate width feature owing to interaction between ejecta-CSM, redshifted broad feature due to interaction with receding part of the ejecta and finally a symmetric component when CSM has engulfed the entire ejecta. We also present two recent observations using 3.6m DOT of SNe~2020cui and 2020rc. Modelling of the H$\alpha$ profile in SN~2020cui shows intermediate features indicates strong ongoing interactions. A broad feature is seen in SN~2020rc due to expanding ejecta which is later overtaken by the ejecta-CSM interaction and leaving behind an intermediate width H$\alpha$ feature. However, viewing angle affects the scenario. Finally, we compare the equivalent width and luminosity of a sample of core-collapse SNe and compare with two SNe~Ibn. This comparison hints that higher luminosity could indicate a more massive progenitor, however, more evidences are required to arrive at a firm conclusions.

%%Appendix

\appendix

%\section{An appendix section}
%Text goes here (Radhakrishnan 1980).
%\begin{equation}
%x=a+b+c
%\end{equation}

%%Use section* for acknowledgements
\section*{Acknowledgements}
		We thank the support staff of the 3.6~m DOT, 2.00~m HCT, 1.30~m DFOT and 1.04~m ST during the observations. We acknowledge Weizmann Interactive Supernova data REPository (WISeREP;\url{http://wiserep.weizmann.ac.il}). KM acknowledges BRICS grant DST/IMRCD/BRICS/Pilotcall/ProFCheap/2017(G). 
		%KM also acknowledge the DST/JSPS grant, DST/INT/JSPS/P/281/2018.
\vspace{-1em}

%%use \balance somewhere in the left column of the last page to balance the two columns in the end page

%%References section

\bibliography{references}
\bibliographystyle{apj}

\end{document}